\documentclass[12pt,dvips]{article}

\usepackage{amsmath,amssymb,exscale}
\usepackage{array,multicol}
\usepackage{afterpage,float,flafter}
\usepackage{epsfig,rotating,pifont}
\usepackage{cite}
\@addtoreset{equation}{section} \makeatother

\def\beq{\begin{eqnarray}}
\def\eeq{\end{eqnarray}}

\def\lsim{\mathrel{\rlap{\lower3pt\hbox{\hskip0pt$\sim$}}
\raise1pt\hbox{$<$}}}         
\def\gsim{\mathrel{\rlap{\lower4pt\hbox{\hskip1pt$\sim$}}
\raise1pt\hbox{$>$}}}         

\title{
\vspace{-3cm}
\begin{flushright}
\small{CERN-PH-TH/2010-088}
\end{flushright}
\vspace{2.7cm}
\huge{Partially Supersymmetric\\ Composite Higgs Models}
\vspace*{0.7cm}
\author{
\Large{\text{Michele Redi}\footnote{michele.redi@cern.ch}~~{\large and} \text{Ben Gripaios}\footnote{ben.gripaios@cern.ch}}\\ \\
\emph{CERN, PH-TH, CH-1211, Geneva 23, Switzerland}
}
}

\date{}
\begin{document}
\maketitle \thispagestyle{empty} \vspace*{-.2cm}

\begin{abstract}
We study the idea of the Higgs as a pseudo-Goldstone boson within the framework
of partial supersymmetry in Randall-Sundrum scenarios and their CFT duals. The Higgs and third generation of the MSSM
are composites arising from a strongly coupled supersymmetric CFT with global symmetry $SO(5)$ spontaneously
broken to $SO(4)$, whilst the light generations and gauge fields are elementary degrees of freedom whose
couplings to the strong sector explicitly break the global symmetry as well as supersymmetry. The presence of supersymmetry
in the strong sector may allow the compositeness scale to be raised to $\sim 10~TeV$ without fine tuning,
consistent with the bounds from precision electro-weak measurements and flavour physics. The supersymmetric flavour 
problem is also solved. At low energies, this scenario reduces to the ``More Minimal Supersymmetric Standard Model''
where only stops, Higgsinos and gauginos are light and within reach of the LHC.
\end{abstract}

\newpage
\renewcommand{\thepage}{\arabic{page}}
\setcounter{page}{1}

\section{Introduction}

An old, but still very attractive, idea is that the Higgs boson is a remnant of some strongly coupled dynamics \cite{oldcomposite}.
This possibility has received new impetus in recent times due to its realization in Randall-Sundrum (RS)
scenarios \cite{rs} and their 4-d CFT duals \cite{rscft}, becoming a realistic alternative to supersymmetry (SUSY)
for the stabilization of the electro-weak scale, see \cite{rsrev} for reviews. Through the mechanism of partial
compositeness and conformal dynamics, these models not only explain the hierarchy between the electro-weak and Planck scales, but also allow the hierarchical flavour structure in the Standard Model (SM) to be generated in an elegant way \cite{flavour}.

Despite these advances, models in which the Higgs is a composite state remain phenomenologically imperfect and require a degree of fine-tuning in order to comply with experimental data. Perhaps the most troublesome bound arises from
precision electro-weak measurements (in particular the $S$ parameter), which require the compositeness scale to be at least $\sim 3\,TeV$ \cite{electroweak}.
Modifications to SM couplings, such as $Z\to b \bar{b}$, also set severe constraints \cite{zbb} and flavour bounds,
in the absence of some suitable mechanism, point to an even higher scale of compositeness \cite{flavourbounds,weiler}. As a consequence
the electro-weak scale is not completely natural and a `little' hierarchy problem still remains.

Recently, building on previous work \cite{partialsusy}, Sundrum has suggested the co-existence of supersymmetry (SUSY) and strong
conformal dynamics in the strong sector, as a way of alleviating the phenomenological bounds on composite Higgs models \cite{emergentsusy}. The starting observation is that a compositeness scale around $10~TeV$, or higher, would relieve most
of the tension from precision electro-weak data and particularly from flavour physics.
Unfortunately, simply raising the strong dynamics scale would introduce a fine tuning of the order of one part {\em per mille}, which is hard to accept in a framework motivated by the hierarchy problem. This can only be avoided if some new physics appears at, or below, the $TeV$ scale
that cancels the quadratic  divergences in the Higgs mass up to the strong dynamics scale. Two mechanisms that realize this scenario may be imagined. One possibility is the Little Higgs mechanism (see \cite{littlereview} for a review), whereby the Higgs is a pseudo Goldstone Boson (GB), protected by multiple global symmetries. This requires, at the very least, a complicated group structure and, even in the most optimistic scenario, it seems difficult to raise the strong dynamics scale as high as $10~TeV$. The second option, on which we focus in this paper, is, of course, supersymmetry.

Merging supersymmetry and strong dynamics, in the framework of partial compositeness
realized in RS scenarios, provides an amusing twist of the standard supersymmetric paradigm.
Indeed, a widespread prejudice in the SUSY literature is that supersymmetric partners of the SM particles should have roughly
the same masses. However, as pointed out in \cite{moreminimalmssm}, a highly non-degenerate spectrum of masses, in which partners of the light generations are heavier than the ones associated to the third generation, would improve the flavour and CP problems in the MSSM, without jeopardizing the solution of the hierarchy problem. The motivation for the ``More Minimal Supersymmetric Standard Model'' (MMSSM) was phenomenological, but it turns out to be  naturally realized in our framework, since partial compositeness
implies partial supersymmetry. In fact, unlike the MMSSM, the superpartners associated to the light SM fermions could be extremely
heavy, solving completely the SUSY flavour problem but without introducing fine-tuning. The light generations are elementary and feel
SUSY breaking strongly, so that the corresponding superpartners are naturally very heavy. By contrast, the Higgs and the top are (mostly)
composites of the strong sector and therefore feel SUSY breaking less and their superpartners can be naturally light. 
Since the Higgs quadratic divergences are, at worst, cut-off by the compositeness scale (which we assume to be around $10~TeV$),  only the third generation, Higgs and gauge superpartners are required below $TeV$ by naturalness\footnote{Since the Yukawa couplings depend on $\tan \beta$ in a supersymmetric theory, in what follows we use ``third generation� as a short hand for ``some or all of the third generation'', depending on their actual couplings.}. Thus, the main phenomenological prediction is that only supersymmetric partners associated to the most important cancellations of the Higgs quadratic divergences will be accessible at the LHC.

In this paper, we elaborate on the framework of partial supersymmetry \cite{partialsusy,emergentsusy},
studying the scenario where the Higgs is a supersymmetric GB, emerging from a strongly-coupled supersymmetric sector
(see also \cite{slava} for a a perturbative realization).
In \cite{emergentsusy}, it was assumed that the supersymmetric Higgs is a composite state of the
strong sector whose mass is anomalously light compared to the dynamical scale. Whilst this is technically
natural in a supersymmetric theory, it nevertheless introduces a  `little' $\mu-$problem. This can be avoided by assuming
that the Higgs is a GB, as, in this case, $\mu$ must vanish in the supersymmetric limit. We investigate this
possibility, effectively generalizing  Ref. \cite{minimalcomposite} to supersymmetry.
Contrary to Ref. \cite{emergentsusy},  the Higgs couples only derivatively to the states of the strong sector and
the Yukawa couplings are obtained through mixing  with states of the elementary sector, which breaks supersymmetry, 
as well as the global symmetries of the strong sector.

As emphasized in \cite{emergentsusy}, one major challenge for this framework comes from the fact that the gauginos, which are states external to the strong dynamics, need to be both light and have supersymmetric couplings, in order to avoid dangerous quadratic divergences in the Higgs mass. Since they are part of the elementary sector, which feels SUSY breaking strongly, neither assumption holds automatically and some model building effort is required. Here, we will focus mostly on the possibility that SUSY breaking in the
elementary sector is large compared to the compositeness scale, but soft (broken by SUSY breaking spurions)
so that couplings are approximately supersymmetric. In this case only the gauginos' mass
must be protected which can be achieved by insisting that the SUSY breaking respects an (approximate) R-symmetry.

Several appealing features of composite Higgs models are automatically inherited. In particular, the mechanism of
partial compositeness operates as in non-supersymmetric scenarios. This allows us to explain the hierarchical structure
of Yukawa matrices, whilst avoiding the most severe flavour bounds associated to the tree level exchange of light resonances.
Let us note that conformal dynamics is needed only to generate hierarchical Yukawa structures, whilst everything else
holds more generally for strongly coupled theories with the same symmetries.
Contrary to the non-supersymmetric scenario, the generation of the Higgs potential is necessarily different in our case. In \cite{minimalcomposite}, the entire potential is generated at one loop from the couplings to the fields of the SM (particularly the top quark), which break the global symmetries of the composite sector. In our scenario, the contribution to the Higgs quartic from the strong sector is small, being cut off by light superpartner masses. Fortunately, supersymmetry automatically
generates a quartic, at tree level, of roughly the right size (though slightly too small) and mass
terms are only generated at one-loop level, once supersymmetry is broken. This allows us to circumvent the fine-tuning problems of
composite Higgs models with a high scale of compositeness.

The outline of the paper is as follows. In Section \ref{4-dmodel}, we describe the features of the strong
sector from which Higgs boson and third generation arise. We explain how this sector is coupled to the
MSSM chiral superfields and how it can be protected by large SUSY breaking effects in the light generations.
In Section \ref{5-dmodel}, using the AdS-CFT correspondence, we translate the physics of the strongly coupled theory into a
five dimensional RS model. This provides useful intuition about the strongly-coupled dynamics, as well
as allowing explicit computations. Going back to the 4-d picture,  we study in Section \ref{potential} the contributions to the Higgs potential due to the couplings to the elementary fields. In Section \ref{scenarios}, we describe various scenarios in which
a successful phenomenology can be achieved. In Section \ref{phenomenology}, we discuss the main phenomenological features of this 
scenario. We conclude in Section \ref{conclusion}.
In two appendices,  we derive the supersymmetric $\sigma-$model corresponding to the coset space $SO(5)/SO(4)$
and introduce a two-site model that economically captures the main features of our setup.

\section{The Four-Dimensional Model}
\label{4-dmodel}

We start by constructing the supersymmetric version of the ``Minimal Composite Higgs Model'' \cite{minimalcomposite}.
The strong sector consists of a strongly coupled large $N$ CFT, with global symmetry $SU(3)\otimes SO(5)\otimes U(1)$.
Here $SU(3)$ corresponds to the colour gauge group and the electroweak group is contained in $SO(5)\otimes U(1)$, the $U(1)$ being  needed to reproduce the correct SM matter hypercharge assignments\footnote{In realistic supersymmetric models the hypercharge should be embedded into an non-abelian group in order to avoid problems with Fayet-Iliopoulos terms \cite{emergentsusy}. The same choices considered there can be used here.}.
In the infrared (IR), the strong dynamics spontaneously breaks $SO(5)$ to $SO(4)\sim SU(2)_L\otimes SU(2)_R$,
delivering GBs in the vector representation of $SO(4)$, corresponding to a complex Higgs doublet in $SU(2)_L$ notation.

In this paper, we will further assume that the strong sector is supersymmetric. Most of what follows will not depend on 
the assumption of conformal dynamics. The spontaneous breaking of the global symmetry 
again delivers four GBs, but now accompanied by a further four scalars, which are massless by supersymmetry.\footnote{In general, given the breaking pattern $G\to H$, the supersymmetric extension
is not uniquely determined and the number of chiral fields can be smaller than the dimension of $G/H$
(see \cite{lerche}). In the next section, we will see that, for
CFTs with holographic duals, the supersymmetric extension always requires the maximal number of chiral fields.}
Such a breaking can be realized by the VEV of a chiral field in the vector representation of $SO(5)$ (see Appendix A).
Using the parametrization of \cite{minimalcomposite}, the supersymmetric GBs are given by
\begin{equation}
\Sigma=f\,\frac {\sin h/f} h \left(h_1,~h_2,~h_3,~h_4,~h~ \cot h/f\right),~~~~~~~~~~~~~~~~~~~~~ h=\sqrt{\sum h_i^2},
\label{goldstone}
\end{equation}
where $h_i$ are chiral superfields and $f$ is the usual decay constant of the $\sigma-$model. Note that $\Sigma$ is, appropriately, a holomorphic function of $h_i$: the real parts correspond to the true GBs whilst the imaginary parts are the SUSY scalar partners. The K\"ahler potential of the supersymmetric $\sigma$-model is simply obtained
from the canonical kinetic term of $\Sigma$,
\begin{equation}
K= \sum_{i=1}^5 \,\Sigma_i(h) \Sigma_i(\bar{h}).
\label{kahler1}
\end{equation}
Interestingly, at two derivative order, other invariants can also be constructed which will be generated by loop corrections.
As we will see in later sections, these could give important contributions to the T-parameter.

Using the mapping between $SO(4)$ and $SU(2)_L\otimes SU(2)_R$, we can make contact with the standard
parametrization of the MSSM. The four chiral fields can be organized in two chiral fields, which are doublets of $SU(2)_L$
and have opposite hypercharge ($T^3_R$), as follows:
\begin{eqnarray}
H_u =\frac 1 {\sqrt{2}} \left(
\begin{array}{c}
h_2 -i h_1  \\
h_4+i h_3
\end{array}
\right),~~~~~~~~~~~~~~~~~~~~
H_d = \frac 1 {\sqrt{2}} \left(
\begin{array}{c}
-h_4 +i h_3  \\
h_2+ i h_1
\end{array}
\right).
\label{higgs}
\end{eqnarray}

The K\"ahler potential (\ref{kahler1}) may then be written as,
\begin{equation}
K=\bar{H}_u H_u+\bar{H}_d H_d -\frac 1 {3 f^2} [H_u H_d+ \bar{H}_u \bar{H}_d](\bar{H}_u H_u+\bar{H}_d H_d)+\frac 1 {f^2}(H_u H_d)(\bar{H}_u \bar{H}_d)+\dots
\label{kahler2}
\end{equation}
The non-renormalizable terms, suppressed by powers of the decay constant, are a distinctive feature of the model; as we shall see, they have important phenomenological consequences. 

We now consider the coupling to the elementary sector, composed of SM fermions and $SU(3)\otimes SU(2) \otimes U(1)$ gauge fields. The basic assumption is that the elementary degrees of freedom couple linearly (to leading order)  to composite operators of the strong sector. These couplings are of two kinds: to currents, associated with global symmetries of the strong sector and to fermionic operators of the strong sector. The first correspond to gauging of the SM $SU(3)\otimes SU(2) \otimes U(1)$ gauge symmetry, whilst the second give rise to the Yukawa couplings for the SM fermions.

Since the SM fields come in incomplete representations of $SO(5)$ and since they do not respect supersymmetry, these couplings explicitly break all the symmetries of the strong sector. For clarity of exposition, let us start the discussion in the limit in which the couplings respect supersymmetry, and then later on add supersymmetry-breaking effects.
We thus couple the MSSM matter and gauge fields to the SCFT. The gauging of the global symmetries is achieved via
\begin{eqnarray}
{\cal L}_{gauge}&=&\int d^2 \theta W_{\alpha}^a W^{\alpha}_a + h.c. + g \int d^4 \theta V_i J_i+\dots\,,  \nonumber \\
J&=&j_0+ i \theta  j_{1/2}-i \bar{\theta} \bar{j}_{1/2} - \theta \sigma^\mu \bar{\theta} j_\mu+ \dots\,,
\label{lgauge}
\end{eqnarray}
where $J_i$ are the $SU(3)\otimes SU(2)_L \otimes U(1)_Y \in SU(3)\otimes SO(4)\otimes U(1)$ supercurrents of the strong sector 
and the ellipsis stands for higher-order couplings to the vector superfields that are required by gauge invariance. 
Note that the supercurrent contains a lowest scalar component, $j_0$, of dimension two. When written in terms of physical fields, the gauging generates a potential for the Higgs from the $D-$term, see also \cite{continonomura}. 
At the renormalizable level this is, of course, the same $D-term$ that arises in the MSSM,
\begin{equation}
V_D=\frac {g^2+g'^2}8 (|H_u|^2-|H_d|^2)^2+\frac {g^2}2 |H_u^\dagger H_d|^2+\dots
\label{dterm}
\end{equation}
with corrections suppressed by powers of $f$. 

The coupling to the strong sector of the MSSM chiral superfields, denoted by $\Phi_i$, and their conjugates, $\Phi_i^c$, 
is given by superpotential couplings to chiral operators of the SCFT, such that
\begin{eqnarray}
{\cal L}_{chiral}&=& \int d^2 \theta \lambda_i \Phi_i O_i^c +\lambda_i^c \Phi_i^c O_i\,, \nonumber \\
O&=&O_0+\theta O_{\frac 1 2} +\theta^2 O_F\,,
\label{lchiral}
\end{eqnarray}
where $O_i$ are chiral operators, characterized by the scaling dimension $\Delta$ of their lowest scalar components.
These couplings introduce a mixing between the elementary states and the composites of the strong sector.
Assuming $m_\rho$ to be the typical scale of the lowest resonances, we have, from large-$N$ counting arguments, that
\begin{equation}
m_\rho \sim \frac {4 \pi f}{\sqrt{N}},
\end{equation}
where $N$ denotes the number of colours of the SCFT. The coupling $g_\rho=4 \pi/\sqrt{N}$
represents the effective expansion parameter of the strongly coupled, large-$N$ theory.
Integrating out the strong sector below $m_\rho$, one obtains supersymmetric Yukawa couplings,
\begin{equation}
y_i \sim \frac {\lambda_i \lambda_i^c}  {g_\rho}\,,
\label{yuk}
\end{equation}
where $\lambda_i$ are evaluated at the IR scale.

The previous equation holds generally in the scenario of partial compositeness, independently of the nature of the strong sector. 
Assuming strong conformal dynamics,  the mechanism of partial compositeness to generate hierarchies between the $\lambda$s,
which in turn control the SM Yukawa couplings, works as in the non-supersymmetric case \cite{minimalcomposite}.
By supersymmetry, the couplings $\lambda_i$ are not renormalized perturbatively, as long as supersymmetry is unbroken.
Nevertheless, the physical couplings depend on the scale, at the classical level, through the wave-function.
A logarithmic dependence for the physical Yukawa couplings also appears from the renormalization of
the K\"ahler potential. Therefore the same RG equation as the one in \cite{minimalcomposite} is obtained.
Through the RG evolution, large hierarchies can be generated, depending on the dimension of the strong-sector operators to which the SM
fields couple. If $\lambda$ in eq. (\ref{lchiral}) is irrelevant in the IR, small Yukawa couplings are generated in the IR, 
as appropriate for the first and second generations, whilst if $\lambda$ is relevant it can 
generate a large Yukawa coupling, as observed for the top quark. Explicitly, one finds
\begin{eqnarray}
\lambda_{IR}&\sim& g_\rho \left(\frac {\mu_{IR}}{\Lambda}\right)^{\Delta-2} \,,~~~~~~~~~~~~~~~~~~~\Delta>2,\nonumber \\
\lambda_{IR}&\sim& g_{\rho}\sqrt{2-\Delta}\,,~~~~~~~~~~~~~~~~~~~~~\Delta<2,
\end{eqnarray}
where $\Lambda$ is the UV cutoff, around the Planck scale.

In our picture, the first two generations will couple to chiral operators of dimension larger than
two so they may be considered elementary. To protect the Higgs from quadratic divergences at the TeV scale, we will need to take
the entire third generation of quarks (or at least the $q_L$ and $t_R$) to be mostly composite. Then, both the left- and right-handed stops remain light. This is necessarily different from non-supersymmetric scenarios with a lower compositeness scale, where consistency with precision electro-weak measurements implies that only the $t_R$
should be mostly composite. 

In the supersymmetric limit, we therefore obtain precisely the MSSM, where the two Higgs doublets are
GBs and their supersymmetric partners, plus a tower of higher dimensional
supersymmetric operators suppressed by powers of the compositeness scale. However, as long as supersymmetry is unbroken
there is no Higgs mass parameter. In particular, whilst the supersymmetric coupling to the external sector breaks the Goldstone symmetry of the Higgs, no superpotential can be generated for $H_u$ and $H_d$ perturbatively, due to the non-renormalization of the superpotential; only a quartic potential arises, from $D-$terms. This double protection is a key feature of our model: a mass is forbidden both by Goldstone symmetry and by supersymmetry. This is both a bonus and a liability: on one hand, the GB nature of the Higgs explains why the $\mu$ term is not around the compositeness scale (as  would otherwise be expected, based on naturalness arguments), but, on the other hand, one must explain how to generate a mass for the Higgsino. 

\subsection{SUSY breaking}

The basic assumption is that the SUSY breaking sector couples mainly to the elementary sector 
represented by the chiral fields $\Phi_i$ whilst the composite sector feels SUSY breaking only or mostly through the mixing.
We parameterize SUSY breaking in the elementary sector by the VEV of a chiral superfield $X= \theta^2 m_0$.
Note that we can assign an appropriate $R-$charge to $X$  so that $R-$symmetry is preserved. The term
\begin{equation}
{\cal L}_{susy~breaking} \sim \, a_{ij} \int d^4 \theta \bar{X}  X\,  \bar{\Phi}_i \Phi_j
\label{susybreaking}
\end{equation}
generates mass terms of order $m_0$ for the elementary scalars (assuming $a_{ij}$ order one numbers). 
For the first two generations, which are mostly elementary, this translates immediately into a mass of order $m_0$ for the scalars. 
We assume $m_0$ to be large with respect to the compositeness scale but smaller than the Planck scale, in order that the effective theory make sense. Then, the scalars of the light generations will be effectively decoupled and the theory will not be
supersymmetric in this sector. On the other hand, SUSY breaking will be felt very weakly by the states which are
mostly composite, since these are protected by the mixing. As a consequence, the third generation superpartners 
can be light, even for a large value of $m_0$. This can be illustrated using a two site model analogous to that of \cite{contino2sites},
which we construct in Appendix \ref{appendixa}. The physical masses depend on the
mixing angles $\varphi_i$ between elementary and composite states. Schematically, one finds
\begin{equation}
m_{ij}^2 \sim a_{ij}\, m_0^2\,  \cos \varphi_i \cos \varphi_j,
\label{softmasses}
\end{equation}
for the scalar masses of $q_L$, $u^c$ $d^c$. No $A-$terms are generated at tree level since Higgs is fully composite. 
Note that the structure is controlled by the mixings which determine the 
Yukawa couplings. Just as for the latter, hierarchies in $m_{ij}$ are generated by the hierarchical mixings. Since $\varphi_{1,2}\sim 0$ and $\varphi_3\sim \pi/2$,
two eigenvalues are of order $m_0$ and one is of order $\sim m_0 \cos\varphi_3$, which we require to be below a TeV. 
The off-diagonal entries,  $m_{i3}^2$, then are at most $\sim m_0 \times TeV$.

If the strong dynamics is conformal, the hierarchies are generated by the RG flow. 
For the diagonal entries, one finds
\begin{equation}
m_i \sim m_0 \left(\frac {\mu_{IR}}{\Lambda}\right)^{2-\Delta_i}
\label{softmasscft}
\end{equation}
and similarly for the others, valid as long as $m< \mu_{IR}$. 

The situation is different for the soft terms in the Higgs sector. Half of the scalar degrees 
of freedom are GBs and they remain massless at tree level even when supersymmetry is broken
explicitly. The remaining half are supersymmetric partners of GBs and therefore can acquire 
a mass at tree level, once supersymmetry is broken. It is easy to see that requiring that no mass term 
is present for the GBs implies the following relations among the usual MSSM soft terms:
\begin{eqnarray}
m_{H_u}&=&m_{H_d},\nonumber \\
B_\mu&=& m_{H_u}^2+\mu^2.
\end{eqnarray}
The scalar partners of the GBs are associated formally to operators of dimension $\Delta=1$. From
eq. (\ref{softmasscft}), their mass is exponentially suppressed, unless the SUSY breaking is maximal. 
At loop level, SUSY breaking is transmitted to the composite sector and scalars 
can acquire masses of the order of the electro-weak scale. 
The Higgsino mass will also be extremely small at tree level,
unless the breaking is maximal. Here loops will not help, due to chiral symmetry, and other contributions will be needed.

We could also write down a term giving a large mass to the gauginos. Such a term can however be forbidden or suppressed by a chiral symmetry (R-symmetry). We will assume this to be the case. This is of course 
a necessary condition in order to have approximate supersymmetry in the Higgs sector.\footnote{We note that any $R$-symmetry will, at best, be approximate, at least in a supergravity framework where cancellation of the vacuum energy necessarily requires $R$ to be broken. Nevertheless, the breaking of $R$ can easily be sufficiently small \cite{giudiceromanino}.}

\section{The Five-Dimensional Construction}
\label{5-dmodel}

In this section we sketch the construction of the five-dimensional, effective theory,
corresponding to the strong sector being a CFT. 
By the standard rules of the AdS-CFT correspondence at large $N$, the physics of a conformal, strongly-coupled theory can be mapped into a five-dimensional, weakly coupled theory on a slice of AdS space. 
Given a specific model,
the weakly coupled nature of the 5-d theory allows detailed computations to be performed and provides 
a useful geometric intuition on this type of models.

We take the AdS metric to be
\begin{equation}
ds^2=\frac {L^2}{z^2}\left({dx^2+dz^2}\right),
\label{metric}
\end{equation}
where $L<z<z_{IR}$. In order to reproduce the hierarchy between the Planck and weak scales, we set $z_{IR}/z_{UV}\sim 10^{16}$.
Again, for clarity of exposition, we discuss first the supersymmetric limit and subsequently introduce supersymmetry-breaking effects.

\subsection{Supersymmetric Action}

Global symmetries of the CFT correspond to gauge symmetries of the 5-d theory, so that
the bulk action should be a supersymmetric $SU(3)\otimes SO(5)\otimes U(1)$ gauge theory.
We find it useful to write the action in 4-d $\mathcal{N}=1$ supersymmetric language \cite{martipomarol}, such that
\begin{eqnarray}
S&=&  \frac 1 {g_5^2} \int \,\left(\frac L z\right)\,d^4 x dz\left[\frac 1 4 \int d^2 \theta\, W^{\alpha}W_{\alpha}+h.c.+ \int d^4\theta \left(\partial_z V-\frac L {\sqrt{2}\,z} (\Sigma+\bar{\Sigma})\right)^2 \right], \nonumber \\
V&=&-\theta \sigma^{\mu} \bar{\theta} A_\mu-i \bar{\theta}^2 \theta \lambda_1+i \theta^2 \bar{\theta}\bar{\lambda}_1+\frac 1 2 \bar{\theta}^2 \theta^2 D, \nonumber \\
\Sigma&=& \frac 1 {\sqrt{2}}(\phi+i A_5)+\sqrt{2} \theta \lambda_2+ \theta^2 F_{\Sigma},
\label{gaugeaction}
\end{eqnarray}
where we neglected non-abelian interactions, which will not play an important role in what follows.
In components, the supersymmetric gauge multiplet includes a Dirac fermion, $(\lambda_1, \lambda_2)$, with mass $1/(2L)$ and a real scalar, $\phi$, whose mass
is at the Breitenlohner-Freedman bound, $m^2=-4/L^2$ (associated with operators of the CFT of dimension
5/2, and 2, respectively). The spontaneous breaking, $SO(5)\to SO(4)$, of the CFT can be realized by imposing 
Dirichlet boundary conditions on the IR brane for the gauge bosons associated with the broken symmetries and Neumann boundary conditions for the gauge bosons of $SO(4)$. The unbroken $\mathcal{N}=1$ supersymmetry then implies analogous boundary conditions for the other components in the chiral multiplet. In order to gauge the standard model gauge group on the UV brane, the gauge fields 
corresponding to $SU(3)\otimes SU(2) \otimes U(1)$ have Neumann BCs
whilst the orthogonal combinations
have Dirichlet BCs.
 Appropriate supersymmetric kinetic terms should also be included
on the UV boundary to reproduce the correct SM gauge couplings. With supersymmetric boundary conditions on the UV brane,
$\Sigma$  delivers a massless $N=1$
chiral supermultiplet in the vector representation of $SO(4)$. Note that, for any pattern of breaking $G\to H$, this construction always
produces a number of superfields equal to the dimension of the coset, whilst a generic supersymmetric $\sigma-$model
could contain fewer, see \cite{lerche}.

Using eq. (\ref{higgs}) in $SU(2)\otimes U(1)_Y$ notation, the scalar components are
\begin{eqnarray}
H_u = \frac 1 {\sqrt{2}}\left(
\begin{array}{c}
\phi_+ +i A_+  \\
\phi_0+i A_0
\end{array}
\right),~~~~~~~~~~~~~~~~~~~~
H_d = \frac 1 {\sqrt{2}}\left(
\begin{array}{c}
-\phi_0^+- i A_0^+  \\
\phi_+^++iA_+^+
\end{array}
\right).
\end{eqnarray}
The boundary conditions for $V$ give rise to the $SU(3)\otimes SU(2)\otimes U(1)$
supersymmetric vector multiplet, whose profile is flat in the extra-dimension.
As for the GBs, the Higgs superfields are
localized in the IR.

Let us now consider the matter sector. In the simplest scenario, each chiral superfield of the MSSM
is embedded in a separate bulk hypermultiplet. For a single bulk field, the action reads, schematically
\begin{eqnarray}
S&=&\int\,\left(\frac L z\right)^3\,  d^5 x \int d^4 \theta \left(\bar{\Phi} e^{-V} \Phi+\Phi^c e^{V} \bar{\Phi}^c\right) \nonumber \\
&+&\int d^2\theta  \left(\frac 1 2 \Phi^c \partial_z \Phi-\frac 1 2 \partial_z \Phi^c  \Phi -\frac L {\sqrt{2}\,z}\Phi^c \Sigma \Phi -\frac c z \Phi^c \Phi \right) +h.c.
\label{hyperaction}
\end{eqnarray}
where $\Phi$ and $\Phi_c$ are chiral fields transforming in conjugate representations of the gauge group. The mass parameter, $c$,
is related to the conformal dimension of the operator \cite{continofermions} via
\begin{equation}
\Delta= 1 +\left|c+\frac 1 2\right|
\end{equation}
By supersymmetry, and consistently with the equations of motion, either $\Phi$ or $\Phi^c$ must vanish at each boundary. Massless modes are obtained by insisting that $\Phi$ (or $\Phi^c$) vanishes on both boundaries. If $\Phi$ vanishes, the normalized wave-functions for the complex scalar and Weyl fermions and are,
\begin{eqnarray}
\phi&= & f(c)\,\frac {z_{IR}}{L^2} \left(\frac z{z_{IR}}\right)^{\frac 3 2-c},\nonumber \\
\chi&= & f(c)\,\,\frac {z_{IR}^{\frac 3 2}}{L^2}  \left(\frac z {z_{IR}}\right)^{2-c},\nonumber \\
f(c)&=& \sqrt{\frac {1-2 c}{1-\left(\frac {z_{IR}}L\right)^{2c-1}}},
\end{eqnarray}
whilst if $\Phi^c$ vanishes, one should replace $c\to -c$. Massless modes with $c>1/2$ are UV localized, whilst those with
$c<1/2$ are IR localized, corresponding to small or large mixing with the composite sector.

To build a concrete model, we assume for simplicity that all the fields are
in the four-dimensional, spinorial representation of $SO(5)$; other choices of representation previously considered in non-supersymmetric models
can also be made\footnote{In the absence of supersymmetry, fermions in the spinor representation
of $SO(5)$ are strongly disfavoured phenomenologically, due to the large corrections to $Z\to b\bar{b}$ that result. As we will see in Section \ref{phenomenology}, raising the compositeness scale to $10~TeV$ renders this choice viable.}. The three MSSM quark chiral superfields
originate from three bulk hypermultiplets $\Psi_Q$, $\Psi_u$ and $\Psi_d$. Following \cite{weiler}, we embed them as
\begin{equation}
\Psi_Q=\left(
\begin{array}{c}  q_q[+,+] \\
u_q^c[-,+]\\ d_q^c[-,+] 
\end{array} \right),~~~~~~~~~~~
\Psi_u= \left(
\begin{array}{c}  q_u[+,-] \\
u_u^c[-,-]\\ d_u^c[+,-]
\end{array} \right),~~~~~~~~~~~
\Psi_d= \left(
\begin{array}{c}  q_d[+,-] \\
u_d^c[+.-]\\ d_d^c[-,-]
\end{array} \right)
\end{equation}
where the left-most (right-most) $\pm$ implies that the component $\Phi_c$ or $\Phi$  of the 5-d superfield vanishes on the UV (IR) brane, respectively.
With these boundary conditions, the massless spectrum reproduces the MSSM, but without Yukawa couplings.

In order to generate supersymmetric Yukawa couplings, we need to introduce mixings between different bulk multiplets.
These correspond to superpotential couplings on the IR brane. Consistently with the $SO(4)$ symmetry we may write
\begin{equation}
{\cal L}_{IR}= \left(\frac L {z_{IR}}\right)^3\int d^2 \theta\, \left[\tilde{m}_u \Phi_{q_q} \Phi_{q_u}^c+ \tilde{m}_d \,\Phi_{q_q}\Phi_{q_d}^c+\tilde{M}_u\,(\Phi_{u_q^c}\Phi_{u_u^c}^c+\Phi_{d_q^c}\Phi_{d_u^c}^c)+\tilde{M}_d\,(\Phi_{u_q^c}\Phi_{u_d^c}^c+\Phi_{d_q^c}\Phi_{d_d^c}^c)\right]
\end{equation} 
These boundary terms and the resultant mixings imply that each chiral field of the MSSM lives in multiple bulk multiplets.
In this way, Yukawa couplings are generated from the bulk covariant derivatives, as
\begin{eqnarray}
\lambda_u&\sim& \frac {g_5}{2\sqrt{2}} f_q (\tilde{m_u}-\tilde{M_u}) f_{-u}, \nonumber \\
\lambda_d&\sim& \frac {g_5}{2\sqrt{2}} f_q (\tilde{m_d}-\tilde{M_u}) f_{-d}.
\end{eqnarray}

We have, therefore, been able to reproduce the MSSM (without SUSY breaking) from an extra-dimension, where the two Higgs doublets are supersymmetric GBs. An important feature, however, is that no $\mu$ term or higher-dimensional Higgs superpotential is generated in the supersymmetric limit, since this is inconsistent with the Goldstone symmetry. 
Nevertheless, even at tree-level there is a potential from the $D$-term (\ref{dterm}), which arises from the non-Abelian part of the bulk gauge kinetic term,
and from SUSY-breaking effects, to which we now turn.

\subsection{SUSY breaking}

The SUSY breaking picture described in Section \ref{4-dmodel} is readily translated into five dimensions. We neglect, for simplicity, the flavour 
structure. In the simplest scenario,
the SUSY breaking sector couples only to the elementary fields which correspond to the UV values of the 5-d fields. 
UV localized, SUSY-breaking terms of the form (\ref{susybreaking}) generate contributions to the boundary masses of the scalars,
lifting their massless zero modes. We assume SUSY breaking to be somewhat larger than the compositeness scale $\sim 1/z_{IR}$.
For the UV localized zero modes ($c>1/2$), associated to the elementary chiral fields of the MSSM, the resulting scalar superpartner masses are of the same order of the soft terms on the UV brane.
The scalar super-partners of the light fermions are then well above the compositeness scale. On the other hand, for IR localized states, corresponding to degrees of freedom which belong mostly to the strong sector, the supersymmetry breaking is felt weakly. In general, with boundary mass $m_0$, the mass of the lightest mode scales as
\begin{equation}
m \sim m_0 \left(\frac {z_{IR}}{z_{UV}}\right)^{c -\frac 1 2},
\label{scalarmass}
\end{equation}
which holds as long as $m \ll 1/z_{IR}$, reproducing eq. (\ref{softmasscft}).
This formula shows that, for $c<1/2$ ($\Delta<2$), SUSY-breaking effects are suppressed 
and vanish as $z_{UV}\to 0$. 
For example, already for $c=0.4$,  one obtains scalar masses of order $TeV$ from $m_0=100~TeV$.

After SUSY breaking, the remaining symmetries on the UV brane permit us to write 
\begin{equation}
\int d^4 \theta  X\bar{X} (\partial_z V- \Sigma-\bar{\Sigma})^2,
\end{equation}
which generates soft terms for the real part of $\Sigma$, but not for the imaginary part.
The scalar $\phi$ corresponds to $c=-1/2$ in the formula above, so that
\begin{equation}
m_\phi \sim m_0 \left(\frac {z_{UV}}{z_{IR}}\right),
\end{equation}
which is an exponentially small mass. Similarly, the Higgsino remains almost massless. (Its mass is, in fact, further suppressed, as a consequence of the fact that both chiralities are localized in the IR.) This introduces a phenomenological problem, in that the Higgsino mass is then protected by an almost-exact, chiral symmetry, which forbids a Higgsino mass even after supersymmetry breaking. 

One cure for this problem would be to remove the massless Higgsino mode, by flipping its UV boundary condition. This is somehwat similar to what is done for the custodial partners of the top right in the non-supersymmetric model \cite{custodians}. 
Another cure might be to introduce other sources of symmetry breaking, or indeed other mediators of the existing SUSY-breaking. We return to this issue in Section \ref{scenarios}.

\section{Structure of the Potential}
\label{potential}

A crucial question is how the correct pattern of electro-weak symmetry breaking can be achieved in our framework.
The potential arising from radiative corrections is substantially more complicated that in the non-supersymmetric case. This is due to the fact that the Higgs is doubly protected by supersymmetry and Goldstone symmetry. In fact, such a protection will only be effective in the top, Higgs and gauge sectors. Given a 5-d model, the potential can be computed along the lines of \cite{minimalcomposite}.
The resulting potential is highly model-dependent, being sensitive to the detailed choice of the 5-d parameters as well as to the discrete
choice of representation. Here, we prefer to focus on the structure of the potential that is model-independent, being fixed by the symmetries of the strong dynamics.  We consider, in this context, the circumstances under which the correct pattern of electroweak breaking can be reasonably achieved. We employ purely 4-d language, using the spurion formalism of \cite{silh} (appropriately adapted to account for the effects of supersymmetry). The structure of the potential is determined by the GB nature of the Higgs and by the couplings that explicitly break the symmetries of the strong sector (in this case, both global symmetries and supersymmetry). The model-independent spurions are the gauge couplings and the mixings introduced in Section \ref{4-dmodel} that are necessary to reproduce the MSSM spectrum.
In addition, there may also be other SUSY-breaking effects in the strong sector. For example, SUSY breaking could be mediated to the strong sector not only by SM fields, but also by other interactions.

We now consider, in detail, the spurions associated with the MSSM fields.
In the supersymmetric limit, there is, of course, no contribution from loops to the Higgs potential, except for the renormalization of the $D-$term. We assume that the MSSM chiral fields couple linearly to operators of the strong sector, as in eq. (\ref{lchiral}). 
By integrating out the strong sector in the supersymmetric limit, we obtain the MSSM Yukawa couplings
well non-linerarities,
\begin{equation}
W=\frac {\lambda_u \lambda_u^c}{g_\rho}\int d^2 \theta \, q H_u u^c  F\left(\frac {H_u H_d} {f^2}\right)+\frac {\lambda_d \lambda_d^c}{g_\rho}\int d^2 \theta \, q H_d d^c  F\left(\frac {H_u H_d}{f^2}\right).
\end{equation}
The function $F$, whose argument is the holomorphic combination $h^2 \equiv H_u H_d$, is determined by the Goldstone nature of the Higgs and by the representations of
the operators to which the SM fields couple. 
For the case where the operators are in the $\mathbf{4}$ of $SO(5)$,
one finds \cite{minimalcomposite},
\begin{equation}
F= \frac {f}{h} \sin \left(\frac {h}{f}\right)~,~~~~~~~~~~~~~~~~~~~~~~~~~~h=\sqrt{H_u H_d}.
\end{equation}
Integrating out the composite dynamics, we also obtain corrections to the kinetic terms of the form
\begin{equation}
\frac {\lambda_i^2}{g_\rho^2} \int d^4 \theta\,  \bar{\Phi}_i\Phi_i \left[\cos\left( \frac {h}{f}\right)+h.c.\right]
-\frac {\lambda_i^{c\,2}}{g_\rho^2} \int d^4 \theta\,  \bar{\Phi}^c_i\Phi^c_i \left[\cos\left( \frac {h}{f}\right)+h.c.\right]
\end{equation}

Once supersymmetry is broken, a potential will be generated at one-loop level, in addition to the $D$-term that is already present at tree level. In the absence of tuning or some accidental cancellation, the contribution to the potential
can be easily estimated using the fact that loops will be cut-off at the relevant supersymmetry breaking scale, {\em i.e.} at the mass of the superpartners entering the loops, assuming that they are lighter than $m_\rho$.
The leading contribution, coming from the Yukawa interactions, can be estimated as
\begin{equation}
V_{Yukawa}\sim  \, N_c    \frac {y_{u,d}^2}{16\pi^2} \times m_s^2 \,|H_{u,d}|^2 \times \hat{V}\left(\frac {H_u H_d}{f^2}\right),
\end{equation}
where $m_s$ is the mass of the relevant SUSY partner. In the absence of cancellations, this generates a mass for the up-type Higgs
\begin{equation}
m_{H_u}^2 \sim N_c \frac {y_t^2 }{16 \pi^2}m_{s}^2.
\end{equation}
An analogous expression is obtained for $m_{H_d}^2$.

Note that loops involving from Yukawa interactions will not generate a $B_\mu$ term.
This follows from the fact that, at the level of dimension-four operators, the Peccei-Quinn symmetry is an accidental symmetry of our model.
This symmetry is not, however, a symmetry of the full lagrangian, being broken by non-linear terms involving the GB fields.
Indeed, the $\sigma-$model lagrangian does not possess Peccei-Quinn symmetry and therefore loops involving the Higgs itself will generate
a general form for the potential. We expect
\begin{equation}
V_{Higgs}\sim \frac {m_s^4} {16\pi^2} V\left(\frac {H_u} f,\frac {H_d}f\right)
\end{equation}
where $m_s$ now corresponds to the Higgsino mass. From this, we extract a contribution to the $B_\mu$ term, given by
\begin{equation}
B_\mu\sim \frac 1 {16 \pi^2} \frac {m_s^4}{f^2}
\label{bmu}
\end{equation}
as well as mass terms for $H_{u,d}$ of similar size.

From the field dependent kinetic terms, we also obtain a term
\begin{equation}
V_{kinetic}\sim N_c \frac {m_{s}^4}{g_{\rho}^2}  \times \frac {\lambda^{(c)2}_{u,d}}{16\pi^2}\times \hat{V}\left(\frac {H_u H_d}{f^2}\right)
\label{vpecceiquinn}
\end{equation}
which gives a contribution similar to (\ref{bmu}).
Note that the $B_\mu$ term becomes suppressed, along with the $\sigma$-model non-linearities, at large values of $f$.
In consequence, if these are the main contributions to $B_\mu$, $f$ should not be too large.

Let us next consider gauge loops. Since gauge superfields are mostly elementary degrees of freedom, external to the strong sector, a careless breaking of
supersymmetry in the elementary sector would remove the gauginos from the low energy spectrum, spoiling the naturalness of the
theory. Indeed, at one-loop order, gauge loops would produce a contribution to the Higgs mass of
\begin{equation}
\delta m^2\sim \frac {g^2}{16\pi^2} m_\rho^2,
\end{equation}
which would already require a tuning at the {\em per cent} level or worse.

Here, we will work under the assumption that the gauginos
remain around the weak scale \cite{emergentsusy}. This can be achieved easily enough via an approximate chiral symmetry, requiring that the
SUSY breaking sector respects an $R$-symmetry; see \cite{giudiceromanino}. Obviously, we should, nevertheless, require that the gauginos acquire a mass at least of the order of the weak scale, in order to avoid experimental bounds.

Even with light gauginos, we must also ensure that the gauge couplings respect supersymmetry to a high degree, 
in order that cancellations of gauge contributions to the Higgs mass persist.
In the next Section, we will distinguish the cases where the SUSY breaking is soft (obtained through SUSY breaking spurions) or hard.
In the former case, on which we focus most of our attention, the  gauge couplings are supersymmetric and supersymmetric cancellations will take place at energies above the gauginos mass. In the latter case, the lightness of gauginos alone is insufficient, since the couplings could deviate, by order one amounts, from their supersymmetric values. However, as argued in \cite{emergentsusy}, this case may still be feasible, since supersymmetric couplings can emerge dynamically though the RG evolution induced by the CFT. 

Having determined the structure of the leading potential, we can extract the loop contributions to the quartic coupling of the Higgs.
The main contribution arises from the top loop,
\begin{equation}
\delta \lambda \sim N_c \frac  {m_s^2}{f^2}\frac {y_{L,R}^2} {16 \pi^2}
\end{equation}
This contribution is smaller than in the non-supersymmetric case, being cut-off at $m_s$ rather than $m_\rho$.
 However, it could  be an important addition to the
tree-level quartic coupling if $f$ were comparable to $m_s$. 
In passing, we note that the quartic thus generated does not include
the correction associated with the running of the of the quartic in the MSSM which will be as in the standard case
allowing the Higgs mass to be larger than $m_z$.

Given that one has additional contributions to the quartic compared to those present in the MSSM, 
one might wonder whether the well-known problem of the lightest Higgs mass in the MSSM could be improved in this framework. To address this, one would of course need to compute explicitly the various contributions, in the context of a concrete model. 

\section{Scenarios}
\label{scenarios}

With the above estimates, we can identify the regions of parameter space where phenomenologically acceptable models can be obtained.
We assume supersymmetry to be broken in the elementary sector at a scale higher than the compositeness scale.  For the light generations, which are mostly elementary, this implies that the theory is non-supersymmetric below $m_\rho$. The corresponding Higgs quadratic divergences are then cut-off by the compositeness scale, so that $m_s$ should be replaced by the strong scale $m_\rho$ in the formulas above. Since the mixing with the strong sector is very small (in order to reproduce the small Yukawa couplings), the contribution to the Higgs potential is unimportant, as long as $m_\rho<20~TeV$.

The situation changes for the third generation. Due to the large value of the top (and possibly bottom) Yukawa coupling, the left-handed quark doublet and the right-handed top superfields need to be substantially mixed with the strong sector. This allows the left- and right-handed stops to remain light despite the fact that supersymmetry is broken at a high scale in the elementary sector. In the language of the 5-d theory, this is clearly understood in terms of the localization of zero modes in the extra-dimension. The light generations are UV localized and feel the supersymmetry breaking directly, whilst the third generation is IR localized and feels the SUSY breaking only through the mixing, or via radiative corrections. In general, to reproduce the large top Yukawa either the left-handed quark doublet and the right-handed
singlet have a large mixing with the strong sector, or the top right is fully composite, whilst the left doublet is elementary. For us, the latter option can be discarded, since, in this case, the left-handed stop would be heavy, such that the top quark contribution to the Higgs divergence is insufficiently cancelled.
Therefore we are led to the first option. Assuming, for simplicity, couplings of similar strength, we have
\begin{equation}
\lambda_t \sim \lambda^c_t \sim \sqrt{g_\rho}.
\label{equalcoupling}
\end{equation}

We can further distinguish two scenarios, characterized by the size of SUSY-breaking effects in the elementary sector, and which we refer to as moderate or maximal breaking.

\subsection{Moderate Breaking}

Phenomenologically, the most conservative possibility is that SUSY is broken in the elementary sector at an intermediate scale larger than $m_\rho$, but much below $M_P$. In this case, the stops can be light, even if they are not maximally mixed with the strong sector, {\em cf.} eq. (\ref{scalarmass}). The gauginos can be light, as discussed above, by means of an approximate chiral symmetry and their couplings will be automatically supersymmetric to a high degree. In this case, emergent supersymmetry need not be invoked and the model is described by a softly-broken supersymmetric lagrangian with highly non-degenerate soft parameters. We will consider the phenomenological consequences of this in the next Section.

If the stops are almost massless at tree level in this scenario, then it is hard to generate a phenomenologically-acceptable potential from 
loops of the SM fields, because, at one-loop level, radiative corrections from the top, cut-off by the stops, will be negligible. 
The stops can acquire a mass around $TeV$, for example through loops of gauginos, but the resulting two-loop contribution to the Higgs potential is then too small. 
It is also conceivable that loops of the light quarks, such as $b_R$, which are  sensitive to the scale $m_\rho$, might generate a sizable potential for a significantly large value of $m_\rho$, given that the actual Yukawa couplings are, {\em a priori}, unknown. We will not pursue this idea further here.

The simplest possibility for generating a suitable Higgs potential
would be to add an additional SUSY breaking sector which couples
to the strong dynamics, or to posit that there are extra messengers which communicate SUSY breaking to the strong sector. Indeed, it is not hard to imagine that the SUSY breaking sector couples differently to the
elementary and strong sectors. For example, the SUSY breaking sector could couple directly to the former and via
messengers to the latter. In particular, the direct coupling could respect an R-symmetry, protecting the gaugino masses,
which could then acquire a mass at the electroweak scale through strong sector effects, since the marginally-coupled gauge multiplets feel the breaking
in both the elementary and strong sectors. Since the Higgs is a supersymmetric GB, if SUSY
breaking is mediated to the strong sector by gauge messengers, the $\mu-$problem of gauge mediation could perhaps be solved
along the lines of \cite{muproblem}. 

A second possibility is simply that the stops acquire a mass around a $TeV$ at tree level. This requires 
a coincidence between the size of supersymmetry breaking in the elementary sector and the amount of top mixing
with the strong sector that we will not be able to explain. Working under this assumption, a realistic potential might be generated, but this depends crucially on the value of $f$. From our estimates in the previous section, if $f$ is around the $TeV$ scale, corresponding to $1\ll g_\rho < 4 \pi$, then the Peccei-Quinn symmetry is sufficiently broken by the non-linearities and a $B_\mu$ term of the required size can be generated. 

For $f\sim TeV$, the coupling of the top with the strong sector, eq. (\ref{equalcoupling}), is of intermediate strength. This means that the top is not maximally mixed with the strong sector. 
As for the bottom, recall that
\begin{equation}
\lambda_b \sim \frac {\lambda_t}{50} \tan \beta.
\end{equation}
Therefore, in the region of moderate $\tan \beta$ ($\le10$), in order to reproduce the small bottom Yukawa coupling, the right-handed bottom should be mostly elementary. Even in the absence of superpartners, the smallness of the Yukawa coupling implies that the contribution to the potential is smaller, or of the same order, as the one produced by the top. 
Note that a mass term for $H_d$ is still generated by (\ref{vpecceiquinn}), which is naturally smaller than the mass of $H_u$. This points in the right direction for obtaining the appropriate electro-weak vacuum structure.

The top loops 
also generate a $\mu$ term, of size
\begin{equation} 
\mu\sim  \frac 1 {16 \pi^2} \lambda_t^2 \frac {m_s^3}{f^2}.
\end{equation}
This value is phenomenologically too small. 
This is much the same problem that arises in conventional gauge mediation scenarios. Namely, $\mu-B_\mu$ are generated at the same order in perturbation theory whilst they have different mass dimensions \cite{muproblem}.
One way to overcome this problem could be to couple the Higgsinos to an external fermion with similar quantum numbers as we discuss later.  
Alternatively, some other source of breaking of the global symmetry should be introduced.
For example, $SO(5)$ could be merely an approximate symmetry of the strong sector, whilst supersymmetry remains exact. 
In a theory where the constituents of the composite states are known, this could be achieved, for example, 
by introducing masses for these constituents \cite{luty}, much like the quark masses in the pion lagrangian. 
This breaking would induce a $\mu-$term, as well as higher-order terms.

\subsection{Maximal Breaking}

A very different scenario is obtained when supersymmetry is broken maximally, since it cannot then be described in terms of
a canonical, softly-broken SUSY theory. In the 5-d picture, this scenario can be realized by imposing
twisted boundary conditions for the fermions and bosons on the UV brane. In order for the stops to remain light, the third generation should now be maximally mixed with the strong sector.\footnote{The left- or right-handed chiralities could also be entirely composites of the strong sector (without mixing), in which case they would not feel any breaking at tree level. Note, however, that a least one chirality must be mixed with the elementary sector, in order to generate Yukawa couplings.} This corresponds to choosing $c\sim -1/2$ in the language of Section \ref{5-dmodel}.
Interestingly, at tree level we now have
\begin{equation}
m_{\tilde{t}} \sim  \sqrt{\frac 2 {\log z_{IR}/z_{UV}}} m_\rho.
\label{stopmass}
\end{equation}
With $z_{UV}/z_{IR}\sim 10^{16}$ this implies that the  stop masses are around one-tenth of strong dynamics scale, {\em i.e.}
around a $TeV$. With these tree-level masses, it is possible to obtain an appropriate scalar potential for
the Higgs from the coupling to the MSSM fields alone. The analysis is similar to the one considered in the previous subsection.

This would necessarily require $f\sim TeV$, since otherwise the approximate  Peccei-Quinn symmetry would forbid a large enough $B_\mu$ term.
Note that if the third generation is maximally mixed with the strong sector, the natural value for the top coupling is $g_\rho$, whilst a small
value of $f$ corresponds to $g_\rho\sim 4 \pi$. However, we do not consider this to be a very severe problem, since it the discrepancy could, for example, arise 
through mildly hierarchical mixings.

The Higgsino mass problem can also be solved by twisting the Higgsino boundary conditions. The mass that one obtains is also
given by the formula (\ref{stopmass}). The dual picture of twisting the boundary condition is however different to that for the stop.
Indeed, contrary to the scalar, the flipped boundary conditions for the fermionic Higgsino cannot be obtained by a sub-Planckian SUSY breaking spurion, localized on the Planck brane.  Changing the boundary condition for the fermion corresponds instead to adding elementary degrees of freedom and coupling them to the bulk fermions via Dirac mass terms. Indeed the flipped boundary condition delivers two Dirac Higgsino masses. In consequence, the Higgsino masses cannot be interpreted as a conventional $\mu-B_\mu$ term. Since the left- and right-handed chiralities have different localizations, their wave functions are not strongly affected by the resultant mixing. The IR localized chirality continues to contribute to the cancellation of quadratic divergences of the Higgs mass due to its self coupling. This suffices to protect the Higgs mass up to $10~TeV$.

Thus, this scenario has several nice features. One is that the SUSY breaking scale in the strong sector (represented, say, by the stop
masses) is automatically linked to the strong dynamics scale, via eq. (\ref{stopmass}). Moreover, since there are no light supersymmetric partners for the
 light SM fermions, bounds from flavour and CP violating processes are completely removed. Finally, the Higgsino mass
problem can be overcome without too much extra structure.

Despite these attractive features, various challenges must be faced. First, since the gaugino feels 
the supersymmetry breaking strongly, it becomes harder to maintain the supersymmetric structure at low energies necessary to avoid fine tuning.
The mass can be easily protected by chiral symmetry, as for the moderate breaking scenario. But in the case of maximal breaking,
the couplings would deviate strongly from their supersymmetric values {\em a priori}, destroying the supersymmetric cancellations. In this case we need the mechanism of emergent SUSY \cite{emergentsusy} to generate approximate supersymmetric
couplings in the IR.  Moreover, we should also make sure that gravity mediated SUSY-breaking effects will not spoil the approximate
supersymmetry of the strong sector.  We leave a more detailed study of these challenges to future work.

\section{Phenomenology}
\label{phenomenology}

To summarize what has been gone thus far, we have shown that, under favorable circumstances, the interplay of supersymmetry and Goldstone symmetry may allow the scale of strong dynamics to be postponed to $\sim 10~TeV$, without substantial fine tuning. In fact, since 
the light generations are elementary, one could even go all the way to $m_\rho\le 20~TeV$, without re-introducing large quadratic divergences.
As we now show, this would have dramatic consequences, essentially removing most of the tensions in composite Higgs models 
that arise from precision electro-weak measurements, as well as from flavour bounds. The hierarchical structure of soft terms would, in turn,
solve the SUSY flavour problem. Interestingly, the phenomenology reduces 
to a large extent to that of the the MMSSM \cite{moreminimalmssm} where only the superpartners 
most relevant to the cancellation of quadratic divergences of the Higgs mass are present near the electro-weak scale.

In composite Higgs models, as in technicolour, the main experimental bound arises from
large contributions, controlled by the scale of the new resonances, to the $S$ parameter. In the 
conventions of \cite{silh}
\begin{equation}
\delta\hat{S}\sim \frac {m_w^2}{m_\rho^2}
\end{equation}
Experimentally this should be less than a few parts {\em per mille}, so it is clear that the contribution of the strong dynamics in our framework is negligible.
There are also loop-suppressed contributions to $S$ from the TeV scale superpartners, but these are negligible.
In comparison, the model of Ref. \cite{minimalcomposite}, with $m_\rho\sim 3~TeV$, is at the experimental bound.

A robust feature of our framework is that both the left- and right-handed top need to be strongly mixed with the strong sector to guarantee
supersymmetric cancellations for the third generation. Depending on $\tan \beta$, the left-handed bottom may also require a large mixing
with the strong sector.  Therefore, one should worry about deviations of the coupling $Z\to b \bar{b}$ from the SM value, which is measured with
a precision of a  few parts {\em per mille}. The correction can be estimated, as in \cite{silh}, to be
\begin{equation}
\frac {\delta g_b} {g_b} \sim  {y_L^2} \frac {v^2}{m_\rho^2},
\end{equation}
which is small, even for maximal coupling $y_L \sim g_\rho$. In particular, contrary to the non-supersymmetric scenario, even the minimal
choice of fermions in the $4$ of $SO(5)$ is compatible with the experimental bound.

The remaining outstanding concern is the $T-$parameter. In non-supersymmetric composite Higgs models, the accidental $SO(4)$ custodial symmetry
of the SM Higgs sector is promoted to an exact symmetry of the strong sector, in order to forbid large, tree-level corrections to $T$.
To wit, the Higgs VEV breaks $SO(4)\to SO(3)$, guaranteeing the relation $m_w^2= m_z^2 \cos \theta$ at tree level.
In the presence of multiple Higgs doublets, $SO(4)$ is, in general, insufficient and extra-measures must be taken \cite{us}.
This is because the two Higgs VEVs will not, in general, be aligned and their combined VEVs will break $SO(4)\to SO(2)$.
Large corrections to $T$ may result. Indeed, using the mapping (\ref{higgs}), one finds that, 
in the minimal supersymmetric case with two Higgs doublets, VEV alignment corresponds to $\tan \beta =1$. 
This is of course ruled out phenomenologically, because the lightest Higgs mass would vanish at tree level. Fortunately, the structure the of
K\"ahler potential (\ref{kahler2}) is special and, in fact, no contributions to $T$ arise at tree
level. Indeed, $T$-violating operators have the form, $(\bar{H}_{u,d}H_{u,d})^2$ and do not appear in (\ref{kahler2}).
(The second term in (\ref{kahler2}) also violates custodial symmetry, but it gives a contribution to the
$W$ masses which is proportional to the leading order one and hence does not contribute to $T$ at tree level.)
This remains true to all orders in the $\sigma-$model of eq. (\ref{kahler1}).
Corrections will be generated by loops, which give a contribution to $T$ within the experimental bound, for $f$ in the $TeV$ range (see appendix A).  Note that, in the case where $B_\mu$ is generated by the non-renormalizable terms of the effective lagrangian, some tension may result from taking $f$ to be larger than a $TeV$. A detailed analysis would be needed in a specific model.

Given the fact that the scale of the new dynamics could be $10~TeV$ or higher, one might also wonder whether
the custodial $SO(4)$ symmetry is necessary in the first place. A generic $\sigma-$model producing a complex Higgs doublet,
such as $SU(3)/SU(2)\otimes U(1)$, will give a contribution to $T$ from the non-linearities \cite{silh} of size
\begin{equation}
\delta T\sim\frac {v^2} {f^2}.
\end{equation}
Since $\delta T\le  10^{-3}$, experimentally this requires $f \ge 7-8~TeV$. For such a large value of $f$,
the Higgs behaves like an elementary state up to the compositeness scale $m_\rho\sim 10\, TeV$. 
Note that the coupling $g_\rho$ of the strong sector is then similar to the one of the top Yukawa, thus blurring
the distinction between the strong sector and the SM. 

For smaller $f$, the presence of higher-dimensional operators suppressed by this scale
could be interesting phenomenologically, since they could help with the supersymmetric fine tuning problem \cite{seiberg,espinosa}. 
These corrections are genuine features of the `Higgs as a SUSY GB' scenario. In particular, already in the supersymmetric
limit, the potential from the $D-$term contains higher dimension operators.
Due to the gauge coupling, these latter effects are quite small. But after SUSY breaking there might, as we saw in the
previous section, be important corrections to the scalar potential in comparison to the MSSM.

Regarding flavour, the situation is greatly ameliorated both in comparison to composite Higgs models and to
supersymmetric models. It is already a remarkable feature of non-supersymmetric composite Higgs models with partial compositeness
that flavour bounds are much less severe than in technicolour models.
The light generations are mostly elementary and, as a consequence, flavour-violating, effective operators, obtained integrating out
the heavy states of the strong sector, are automatically suppressed by the small mixing with the composite sector.  
This so-called `RS-GIM mechanism' is, however, imperfect and a tension with flavour bounds remains, if the scale of new resonances
is around $3~TeV$, the lowest value allowed by precision electro-weak measurements; see, for example, \cite{weiler}.
With the scale of the new resonances around $10~TeV$, these bounds are presumably greatly reduced, though a full analysis
is required. 

The supersymmetric flavour problem is also improved, because the partners of the light generations 
are automatically heavy. For example the contribution to Kaon mixing,
\begin{equation}
\frac {g^4}{16\pi^2} \frac{(\Delta m_{sd}^2)^2}{m_{\tilde{q}}^6}
\end{equation}
is suppressed and similarly for other processes. Indeed, this was the motivation in \cite{moreminimalmssm} for considering non-degenerate scalar masses. In the MMSSM, the squark masses of the first two generations need to be lighter than $20~TeV$ for the model to remain natural whilst FCNC or $CP$ violating processes would require their masses to be at least $50~TeV$ or $500~TeV$, respectively, so that a certain tension remains. In our scenario, the superpartners of the light generations could be much heavier without re-introducing fine tuning, since the Higgs quadratic divergences are, in any case, cut-off at the scale of the strong dynamics. This allows us to completely decouple the SUSY flavour problem. 

An interesting feature is that, assuming anarchic SUSY breaking in the elementary sector, the structure of the soft term, eq. (\ref{softmasses}), is controlled by the mixings between the elementary and composite sectors. If $m_0$ is sufficiently small, this could produce visible effects; see \cite{giudicehierarchical}. 

Concerning experimental signatures, the resonance states associated
with the strong dynamics, already very challenging for $m_\rho \sim 3~TeV$, become completely inaccessible at the LHC. Anomalous couplings of the Higgs will also be invisible, since higher-dimensional operators involving the Higgs are suppressed by the scale $f$, which is at least a $TeV$. In consequence, the main features of compositeness will have to await precision measurements at a future collider. The only exception could be the custodial partners of the third generation quarks \cite{custodians}. If the top is maximally localized in the IR, the masses of these states (bosonic and fermionic) could be roughly one-tenth of the strong dynamics scale (much like in eq. (\ref{stopmass})) and could be studied at the LHC \cite{partnertopr}.
The main phenomenological prediction will be that only the supersymmetric partners most directly associated to the cancellation of
the quadratic divergences, namely the third generation squarks, together with gauginos and Higgsinos, will be produced at the LHC.

Finally we would like to comment on the fine-tuning in this class of models.
When the potential for the GBs is generated at one-loop level, as in \cite{minimalcomposite}, naturally $H\sim f$ or zero in the absence of cancellations. The amount of fine tuning can then be quantified as the ratio $v^2/f^2$. In the supersymmetric case, we already have a tree-level quartic potential and the appropriate quantity to consider is then the radiative stability of the Higgs mass. The analysis of fine tuning goes through as in the MSSM and requires superpartners to be lighter than a $TeV$ to avoid sizable fine tuning.
Let us note that, since there is no running for the  Higgs mass above $\sim 10~TeV$, the fine tuning will be
reduced compared to the MSSM valid up to the GUT or Planck scale.

\section{Conclusions}
\label{conclusion}

In this paper, motivated by the tension of composite Higgs models with experimental bounds, 
we constructed a Minimal Supersymmetric Composite Higgs Model.  
The rationale is that assuming the strong sector to be supersymmetric may allow the
dynamical scale of the strong sector to be postponed to $10-20~TeV$, whilst retaining naturalness of the electroweak scale. 
This would satisfy all bounds on compositeness. As in the non-supersymmetric scenario, 
through the mechanism of partial compositeness, conformal dynamics allow us to generate Yukawa coupling hierarchies 
from anarchical flavour structures, but other features of the model differ, particularly the generation of the scalar potential for the Higgs.

Perhaps the most remarkable outcome of merging supersymmetry and partial compositeness is that it can 
naturally realize the ``More Minimal Supersymmetric Standard Model'' \cite{moreminimalmssm}, where only
the third generation scalar super-partners are light. In comparison to that scenario, the partners of the first two generations
could be very heavy, without reintroducing fine tuning in the Higgs mass. This would, in turn, solve 
the supersymmetric flavour problem. We find this an intriguing connection that may be worth exploring further. 
The experimental predictions are quite sharp,  namely that only the stops, 
gauginos and higgsinos are light and within reach of the LHC. This constitutes a radical departure 
from the standard expectations in supersymmetric models.

In this work, we did not investigate a detailed model, since many features are model dependent. Nevertheless, it would
be interesting to construct an explicit model with these features and to compare with existing data.
Several scenarios are possible, depending on the amount of supersymmetry breaking in the elementary sector and 
we considered the region of parameters where the phenomenology seemed to be most promising. 
Generating the required potential for the Higgs appears challenging 
and may require extra-sources of breaking of global symmetries or supersymmetry (or the mediation thereof), besides the couplings
to the elementary sector. Also, the relation between the stop masses and the compositeness scale does not follow
automatically. We leave these open problems for future work.

\vspace{0.5cm} {\bf Acknowledgments:}
We are grateful to Christophe Grojean, Giovanni Villadoro, Andreas Weiler and especially Roberto Contino for useful discussions. 

\vspace{-.5cm}

\appendix
\section{Supersymmetric $\sigma-$models}
\label{appendixb}

In order to construct the two-derivative effective action for the Goldstone bosons it is convenient to
consider a linear $\sigma-$model that realizes the breaking. The simplest choice is,
\begin{equation}
{\cal L}=\sum_{i=1}^5\,\left\{\int d^4 \theta\, \bar{\Sigma}_i \Sigma_i+\int d^2 \theta \, \left[ m\, \Sigma_i \Sigma_i - \frac m {2 f^2} \,(\Sigma_i \Sigma_i)^2\right]+\,h.c.\right\}
\label{linearsigma}
\end{equation}
The superpotential is invariant under complexified $SO(5)$ rotations whilst the D-term is invariant under
$SU(5)$ transformations. As a consequence the full action is $SO(5)$ invariant. The vacuum manifold
consists of,
\begin{eqnarray}
\Sigma_i&=&0 \nonumber \\
\Sigma^2&=&f^2
\end{eqnarray}
The first solution corresponds to the unbroken symmetry. The solutions to the second equation
can be parameterized as,
\begin{equation}
\Sigma=f\,\frac {\sin h/f} h \left(h_1,~h_2,~h_3,~h_4,~h~ \cot h/f\right),~~~~~~~~~~~~~~~~~~~~~ h=\sqrt{\sum h_i^2}
\end{equation}
which breaks $SO(5)\to SO(4)$.

Plugging into (\ref{linearsigma}) we obtain the K\"ahler potential for the supersymmetric $\sigma-$model, eq. (\ref{kahler1}). This term does not contribute to the $T$-parameter
despite the fact that it violates custodial symmetry. Note that any polynomial $\bar{\Sigma}_i\Sigma_i$ can be added to the effective action consistently with the symmetries.  Contrary to the non supersymmetric case this generates new terms in the 2 derivative effective action which is therefore not uniquely determined by the breaking pattern $G\to H$. Such terms can be obtained by higher order terms in the kinetic term of
the linear $\sigma-$model above suppressed by the appropriate cut-off. They will also be induced by loops,
\begin{equation}
K=f^2 \left[ \bar{\Sigma}_i\Sigma_i+ \frac {g_\rho^2}{16 \pi^2} (\bar{\Sigma}_i\Sigma_i)^2+\dots\right]
\end{equation}

It is easy to see that the new terms contribute to the $T-$parameter which arises from the structure $(\bar{h}_i h_i)^2$
not present in the leading order effective action.  We find that the contribution to $T$ scales as,
\begin{equation}
\delta T\sim  \frac {g_\rho^2}{16 \pi^2} \left(\frac v f\right)^2
\end{equation}
Assuming $m_\rho \sim 10~TeV$ this will be within the experimental bound for $f\sim 2-3~TeV$.

\section{Two Sectors Model}
\label{appendixa}

In order to derive phenomenological consequences of our framework it is useful to construct an effective 
lagrangian that captures the basic features of partial compositeness. The formalism 
of \cite{contino2sites} can be straightforwardly extended to the supersymmetric scenario and we refer to that paper for all the details. 
The composite sector is replaced by massive chiral fields whose mass corresponds roughly to the first resonance of the strong sector, $M$. The elementary sector is composed by chiral fields with the MSSM quantum numbers which mixes with the heavy states.
We focus on the quark sector the extension to lepton being trivial.

The physical MSSM fields are an orthogonal rotation, induced by the mixing, between the massless and heavy states,
\begin{eqnarray} 
\left(\begin{array}{c}
\phi_i \\
\Phi_i
\end{array}\right)=\left(\begin{array}{cc}
\cos \varphi_i & \sin \varphi_i \\
- \sin \varphi_i & \cos \varphi_i
\end{array}\right)
\left(\begin{array}{c}
\phi^{el}_i \\
\Phi^{co}_i
\end{array}\right)
\end{eqnarray}
These rotations determine the Yukawa's \cite{contino2sites},
\begin{equation}
y_{ij}^u= \sin \varphi_{q_{Li}} \left(Y_*^U\right)_{ij} \sin \varphi_{u_{Rj}}\,,~~~~~~~~~~~~~~~~~~~~~y_{ij}^d= \sin \varphi_{q_{Li}} \left(Y_*^D\right)_{ij} \sin \varphi_{d_{Rj}}
\end{equation}
where $Y_*$ are matrices with order one entries assuming complete anarchy.

In the elementary-composite base the relevant squark mass matrix reads,
\begin{eqnarray} 
\left(\begin{array}{cc}
M^2 \sin^2 \varphi_i & M^2 \sin \varphi_i \cos \varphi_i \\
M^2 \sin \varphi_i \cos \varphi_i & M^2 \cos^2 \varphi_i
\end{array}\right)
\end{eqnarray}
whose zero eigenvalues are associated to the MSSM squarks.

SUSY breaking mass terms for the elementary fields correspond to adding $m_0^2$ in the 11 element of the above matrix.
In limit $m_0 \cos \varphi>> M$ or $m_0 \cos \varphi<< M$ relevant respectively for the light 
and  third generation one finds that the mass eigenstates are approximately the same as in the SUSY limit. The mass of
the supersymmetric partners is then,
\begin{equation}
m_i\sim m_0 \cos \varphi_i
\end{equation}

The previous discussion can be easily extended to the case of non trivial flavour structure.
For the squark masses one finds,
\begin{eqnarray}
\left(m_{\tilde{q}}^2\right)_{ij}= \left(m_{0\,q}^2\right)_{ij} \cos \varphi_{q_{Li}}\cos \varphi_{q_{Lj}}\nonumber \\
\left(m_{\tilde{u}}^2\right)_{ij}= \left(m_{0\,u}^2\right)_{ij} \cos \varphi_{q_{ui}}\cos \varphi_{q_{uj}}\nonumber \\
\left(m_{\tilde{d}}^2\right)_{ij}= \left(m_{0\,d}^2\right)_{ij} \cos \varphi_{q_{di}}\cos \varphi_{q_{dj}}
\end{eqnarray}
and similarly for the leptons. Since the Higgs is part of the composite states and supersymmetry is broken in the elementary
sector, $A-$terms are not generated at tree level. 
Assuming an anarchic structure of the mass terms in the elementary sector  a hierarchical structure
for the soft terms is generated by the mixings. Since the mixings are determined by the Yukawas we obtain a correlation 
between these couplings and the soft terms.

\vspace{-.5cm}


\end{document}